\input amstex
\documentstyle{jams}
\NoBlackBoxes

\define\emph{\it} \define\demph{\it}
\define\BO{{\Cal O}}
\refstyle{A}
\widestnumber\key{DGPR84}

\topmatter
\title Two heads are better than two tapes\endtitle
\rightheadtext{TWO HEADS ARE BETTER THAN TWO TAPES}
\author Tao Jiang, Joel I. Seiferas, and Paul M. B. Vit\'anyi\endauthor
\leftheadtext{TAO JIANG, JOEL I. SEIFERAS, AND PAUL M. B. VIT\'ANYI}

\address
Department of Computer Science,
McMaster University,
Hamilton, Ontario L8S 4K1,
Canada
\endaddress

\email jiang\@maccs.mcmaster.ca\endemail

\address
Computer Science Department,
University of Rochester,
Rochester, New York 14627-0226,
U. S. A.
\endaddress

\email joel\@cs.rochester.edu\endemail

\address
Centre for Mathematics and Computer Science (CWI),
Kruislaan 413,
1098 SJ Amsterdam,
The Netherlands
\endaddress

\email paulv\@cwi.nl\endemail



     \abstract We show that a Turing machine with two single-head
one-di\-men\-sion\-al tapes cannot recognize the set $$\{\, x2x' \mid 
\text{$x \in \{0,1\}^*$ and $x'$ is a prefix of $x$}\,\}$$ in real time, 
although it can do so with three tapes, two two-dimensional tapes, or 
one two-head tape, or in linear time with just one tape. In particular, 
this settles the longstanding conjecture that a two-head Turing machine 
can recognize more languages in real time if its heads are on the {\emph 
same\/} one-dimensional tape than if they are on {\emph separate\/} 
one-dimensional tapes.\endabstract

     \thanks The first author was supported in part by NSERC Operating
Grant OGP0046613.  The third author was supported in part by the
European Union through NeuroCOLT ESPRIT Working Group Number 8556, and
by NWO through NFI Project ALADDIN under Contract number NF
62-376.\endthanks

     \thanks We thank Wolfgang Maass for discussions that contributed to 
an early version of the Anti-Holography Lemma in 1985.  We thank Ming Li 
for other valuable discussions, and for first bringing the problem 
addressed here to the attention of the first author.  We thank Zvi Galil 
and Ken Regan for helpful comments on the manuscript.\endthanks

     \thanks An earlier version of this report appeared in the
Proceedings of STOC '94, the Twenty-Sixth Annual ACM Symposium on the 
Theory of Computing, pp.~668--675.\endthanks

\date shortly after July 20, 1995\enddate


\endtopmatter

\document

\head 1.  Introduction \endhead
     The Turing machines commonly used and studied in computer science 
have separate tapes for input/output and for storage, so that we can 
conveniently study both storage as a dynamic resource and the more 
complex storage structures required for efficient implementation of 
practical algorithms \cite{HS65}.  Early researchers \cite{MRF67} asked
specifically whether two-head storage is more powerful if both heads are 
on the {\emph same\/} one-dimensional storage tape than if they are on 
{\emph separate\/} one-dimensional tapes, an issue of whether {\emph 
shared\/} sequential storage is more powerful than {\emph separate\/} 
sequential storage.  Our result settles the longstanding conjecture that 
it {\emph is}.

     In a broader context, there are a number of natural structural
parameters for the storage tapes of a Turing machine.  These include the 
number of tapes, the dimension of the tapes, and the number of heads on 
each tape.  It is natural to conjecture that a deficiency in {\emph 
any\/} such parameter is significant and cannot be fully compensated for 
by advantages in the others.  For the most part, this has indeed turned 
out to be the case, although the proofs have been disproportionately 
difficult \cite{Ra63, He66, Gr77, Aa74, PSS81, Pa82, DGPR84, Ma85, LV88, 
LLV92, MSST93, PSSN90}.

     The case of deficiency in the number of heads allowed on each tape 
has turned out to be the most delicate, because it involves a surprise:  
A larger number of single-head tapes {\emph can\/} compensate for the 
absence of multihead tapes \cite{MRF67, FMR72, LS81}. For example, four 
single-head tapes suffice for general simulation of a two-head tape 
unit, without any time loss at all \cite{LS81}.  The remaining question 
is just what, if anything, {\emph is\/} the advantage of multihead 
tapes.

     The simplest version of the question is whether a two-head tape is 
more powerful than two single-head tapes.  In the case of {\emph 
multidimensional\/} ``tapes'', Paul has shown that it is \cite{Pa84}.  
His proof involves using the two-head tape to write, and occasionally to
retrieve parts of, algorithmically incompressible bit patterns. Because 
the diameter of the pattern (and hence the retrieval times) can be kept 
much smaller than its volume, no fast simulator would ever have time to 
perform any significant revision or copying of its representation of the 
bit pattern.  On ordinary {\emph one\/}-dimensional tapes, however, 
retrievals take time that is {\emph not\/} small compared to the volume 
of data, and we cannot so easily focus on a nearly static representation 
of the data.  We need some more subtle way to rule out all (possibly 
very obscure) copying methods that a two-tape machine might employ to 
keep up with its mission of fast simulation.  Our argument below does 
finally get a handle on this elusive ``copying'' issue, making use of a 
lemma formulated more than ten years ago with this goal already in mind 
\cite{Vi84\rm, Far-Out Lemma below}.

     Our specific result is that no Turing machine with just two
single-head one-dimensional storage tapes can recognize the following
language in real time:\footnote{{\demph On-line recognition\/} requires 
a verdict for each input prefix before the next input symbol is read, 
and {\demph real-time recognition\/} is on-line recognition with some 
constant delay bound on the number of steps between the reading of 
successive input symbols.  Note that even a {\emph single\/}-tape Turing 
machine can recognize $L$ on-line in {\emph cumulative\/} linear time; 
but this involves an unbounded (linear-time) delay to ``rewind'' after 
reading the symbol $2$.  In cumulative linear time, in fact, {\emph 
general\/} on-line simulation of a two-head one-dimensional tape is 
possible using just two single-head tapes \cite{St70}; so real time is a 
stronger notion of ``without time loss''.  (There is an analogous 
linear-time simulation for two-dimensional tapes \cite{ST89}, but the 
question is open for higher dimensions.)}
$$L=\{\, x2x' \mid \text{$x \in \{0,1\}^*$ and $x'$ is a prefix of $x$}\,\}.$$
With a two-head tape, a Turing machine can easily recognize $L$ in real 
time.

     Our result incidentally gives us a tight bound on the number of
single-head tapes needed to recognize the particular language $L$ in real 
time, since three {\emph do\/} suffice \cite{MRF67, FMR72}.  Thus $L$ is 
another example of a language with ``number-of-tapes complexity'' $3$, 
rather different from the one first given by Aanderaa \cite{Aa74, PSS81}. 
(For the latter, even a two-head tape, even if enhanced by instantaneous
head-to-head jumps and allowed to operate probabilistically, was not
enough \cite{PSSN90}.)

     Historically, multihead tapes were introduced in Hartmanis and
Stearns' seminal paper \cite{HS65}, which outlined a {\emph
linear\/}-time\adjustfootnotemark{-1}\footnotemark\ simulation of an
$h$-head tape, using some larger number of ordinary single-head tapes.
Sto\ss~\cite{St70} later reduced the number of single-head tapes to just 
$h$.  Noting the existence of an easy {\emph real\/}-time simulation in 
the {\emph other\/} direction, Be\v cv\' a\v r \cite{Be65} explicitly 
raised the question of real-time simulation of an $h$-head tape using 
only single-head tapes.  Meyer, Rosenberg, and Fischer devised the first 
such simulation \cite{MRF67}; and others later reduced the number of 
tapes \cite{FMR72, Be74, LS81}, ultimately to just $4h-4$.  We are the
first to show that {\emph this\/} number {\emph cannot\/} always be 
reduced to just $h$, although both the extra power of multihead tapes 
and the more-than-two-tape complexity of the particular language~$L$ 
have been longstanding conjectures \cite{FMR72, LS81, Vi84, Pa84}.

\head 2.  Tools \endhead
\subhead Overlap \endsubhead
     Part of our strategy will be to find within any computation a
sufficiently long {\emph sub\/}computation that is sufficiently well
behaved for the rest of our analysis.  The behavior we seek involves
limitations on repeated access to storage locations, which we call 
``overlap'' \cite{Aa74, PSSN90}.

     Our overlap lemma is purely combinatorial, and does not depend at
all on the {\emph nature\/} of our computations or the ``storage
locations'' corresponding to their steps.  Nor does it depend on the
computational significance of the steps designated as ``distinguished''. 
 The use of computational terminology would only obscure the lemma's 
formulation and proof, so we avoid it.

     An {\demph overlap event\/} in a sequence $S =
\ell_1,\ldots,\ell_T$ (of ``storage locations'', in our application) is a 
pair $(i,j)$ of indices with $1 \le i < j \le T$ and $\ell_i = \ell_j 
\notin \{\ell_{i+1},\ldots,\ell_{j-1}\}$ (``visit and soonest revisit'').  
If $\omega_t(S)$ is the number of such overlap events ``straddling'' $t$ 
(i.e., with $i \le t$ but $j \nleq t$), then the sequence's {\demph 
internal overlap\/}, $\omega(S)$, is $\max\{\,\omega_t(S) \mid 1 \le t < 
T\,\}$. The {\demph relative\/} internal overlap is $\omega(S)/T$.

     Here is an example:  In the sequence
$$S = \text{cow}, \text{pig}, \text{horse}, \text{pig}, \text{sheep}, 
  \text{horse}, \text{pig},$$
the overlap events are $(2,4)$, $(4,7)$, and $(3,6)$.  For $t$ from $1$ 
up to $6$, the respective values of $\omega_t(S)$ are $0$, $1$, $2$, $2$, 
$2$, and $1$; so $\omega(S)$ is $2$, and the relative internal overlap is 
$2/7$.

     (In our setting below, we apply these definitions to the sequence of
storage locations shifted to on the successive steps of a computation or 
subcomputation.  Without loss of generality, we assume that a multihead 
or multitape machine shifts exactly one head on each step.)

     The lemma we now formulate guarantees the existence of a contiguous 
subsequence that has ``small'' relative internal overlap (quantified 
using $\varepsilon$), but that is itself still ``long'' (quantified using 
$\varepsilon'$).  The lemma additionally guarantees that the subsequence 
can include a quite fair share of a set of ``distinguished positions'' of 
our choice in the original sequence.

     (The ``designated positions'' in our setting will be the items in 
the sequence that correspond to a large ``matching''---a notion we define 
later, especially motivated by computations involving two heads.)

     \proclaim{Overlap Lemma}  Consider any $\delta<1$ and any
$\varepsilon>0$.  Every sequence $S$ (of length~$T$, say) with
``distinguished-position'' density at least $\delta$ has a long
contiguous subsequence, of length at least $\varepsilon' T$ for some
constant $\varepsilon' > 0$ that depends only on $\delta$
and~$\varepsilon$, with distinguished-position density still at least
$\delta/2$, and with relative internal overlap less than
$\varepsilon$.\endproclaim

     \demo{Proof}  Without loss of generality, assume $T$ is a power of 
$2$ that is large in terms of $\delta$ and $\varepsilon$.  (If $T$ is 
{\emph not\/} a power of $2$, then we can discard an appropriate prefix 
and suffix of combined length less than half the total, to obtain such a 
sequence with distinguished-position density still at least $\delta$.)  
We consider only the sequence's two halves, four quarters, eight eighths, 
etc.  Of these, we seek many with sufficient distinguished-position 
density (at least $\delta/2$) and with internal overlap accounted for by 
distinct overlap events, planning then to use the fact that each item in 
$S$ can serve as the second component of at most one overlap event.

     Within each candidate subsequence $S'$, we can select a particular 
straddle point~$t$ for which $\omega(S') = \omega_t(S')$, and then we 
can designate the $\omega(S')$ overlap events within~$S'$ that straddle 
position~$t$ as the ones we consider counting.  The designated overlap 
events in $S'$ can be shared by another interval only if that interval 
includes the corresponding selected straddle point~$t$.

     We consider the candidate sequences in order of decreasing length 
(i.e., halves, then quarters, then eighths, etc.). At each partitioning 
level, at least fraction $\delta/2$ of the subsequences must have 
distinguished-position density at least $\delta/2$.  (Otherwise, we 
cannot possibly have the guaranteed total $\delta T$ distinguished 
positions in the subsequences on that level, since $(\delta/2)\cdot1 + 
(1-\delta/2)\cdot\delta/2 < \delta$.)  Among these, we can count 
distinct overlap from
\roster
\item"" $\lceil (\delta/2)2 \rceil = \lceil \delta \rceil \ge \delta/2 - 1/2$ halves,
\item"" $\lceil (\delta/2)4 \rceil - \lceil (\delta/2)2 \rceil = \lceil 2\delta \rceil - \lceil \delta \rceil \ge \delta - 1/2$ quarters,
\item"" $\lceil (\delta/2)8 \rceil - \lceil (\delta/2)4 \rceil = \lceil 4\delta \rceil - \lceil 2\delta \rceil \ge 2\delta - 1/2$ eighths,
\item"" $\lceil (\delta/2)16 \rceil - \lceil (\delta/2)8 \rceil = \lceil 8\delta \rceil - \lceil 4\delta \rceil \ge 4\delta - 1/2$ sixteenths,
\item"" etc.
\endroster
Unless we find one of these sequences that has relative internal overlap 
less than~$\varepsilon$, this accounts, at the $i$th level, for at least
$$(2^{i-2} \delta - t)(\varepsilon T/2^i) = \varepsilon\delta T/4 - \varepsilon T/2^{i+1}$$
distinct overlap events, and hence for more than $T$ distinct overlap 
events after $\lceil (4+2\varepsilon)/(\varepsilon\delta) \rceil$ levels.  
This is impossible, so we must find the desired low-overlap sequence at 
one of these levels.\qed\enddemo

\subhead Kolmogorov Complexity \endsubhead
     A key to the tractability of our arguments (and most of the recent 
ones we have cited \cite{Pa82, Pa84, PSS81, DGPR84, Ma85, LV88, LLV92, 
PSSN90, Vi84}) is the use of ``incompressible data''. Input strings that 
involve such data tend to be the hardest and least subject to special 
handling.

     We define incompressibility in terms of Kolmogorov's robust notion 
of descriptional complexity \cite{Ko65}.  Informally, the Kolmogorov 
complexity $K(x)$ of a binary string~$x$ is the length of the shortest 
binary program (for a fixed reference universal machine) that prints $x$ 
as its only output and then halts. A string $x$ is {\demph 
incompressible\/} if $K(x)$ is at least $|x|$, the approximate length of 
a program that simply includes all of $x$ literally.  Similarly, a 
string $x$ is ``{\emph nearly\/}'' incompressible if $K(x)$ is ``almost 
as large as'' $|x|$.

     The appropriate standard for ``almost as large'' above can depend 
on the context, a typical choice being ``$K(x) \ge |x| - \BO(\log
|x|)$''.  The latter implicitly involves some constant, however, the 
careful choice of which might be an additional source of confusion in our 
many-parameter context.  A less typical but more absolute standard such 
as ``$K(x) \ge |x| - \sqrt{|x|}$'' completely avoids the introduction of 
yet another constant.

     Similarly, the {\emph conditional\/} Kolmogorov complexity of $x$ 
with respect to $y$, denoted by $K(x|y)$, is the length of the shortest 
program that, {\emph with extra information $y$}, prints $x$.  And a 
string $x$ is incompressible or nearly incompressible {\emph relative to 
$y$} if $K(x|y)$ is large in the appropriate sense.  If, at the opposite 
extreme, $K(x|y)$ is so small that $|x| - K(x|y)$ is ``almost as large 
as'' $|x|$, then we say that $y$ {\demph codes\/} $x$ \cite{CTPR85}.

     There are a few well-known facts about these notions that we will 
use freely, sometimes only implicitly.  Proofs and elaboration, when 
they are not sufficiently obvious, can be found in the literature 
\cite{{\rm especially} LV93}.  The simplest is that, both absolutely and 
relative to any fixed string $y$, there are incompressible strings of 
every length, and that {\emph most\/} strings are nearly incompressible, 
by {\emph any\/} standard. Another easy one is that significantly long 
subwords of an incompressible string are themselves nearly 
incompressible, even relative to the rest of the string.  More striking 
is Kolmogorov and Levin's ``symmetry of information'' \cite{ZL70}:  
$K(x)-K(x|y)$ is very nearly equal to $K(y)-K(y|x)$ (up to an additive 
term that is logarithmic in the Kolmogorov complexity of the binary 
encoding of the pair $(x,y)$); i.e., $y$ is always approximately as 
helpful in describing $x$ as vice versa!  (Admittedly, the word 
``helpful'' can be misleading here---the result says nothing at all 
about the relative {\emph computational\/} complexity of generating the 
two strings from each other.)  All these facts can be relativized or 
further relativized; for example, symmetry of information also holds in 
the presence of help from any fixed string $z$:
$$K(x \bigm| z) - K(x|y \bigm| z) \approx K(y \bigm| z) - K(y|x \bigm| z).$$

\head 3.  Strategy \endhead
     For the sake of argument, suppose some two-tape Turing machine $M$ 
does recognize $\{\, x2x' \mid \text{$x \in \{0,1\}^*$ and $x'$ is a 
prefix of $x$}\,\}$ in real time.  Once a binary string $x \in
\{0,1\}^*$ has been read by $M$, the contents of $M$'s tapes tend to
serve as {\emph a very redundant representation of prefixes of $x$},
because $M$ has to be prepared to retrieve them at any time. (Our
problem and this observation were motivation for Chung, Tarjan, Paul,
and Reischuk's investigation of ``robust codings of strings by pairs of 
strings'' \cite{CTPR85}.) One way around this is for $M$ to keep one or 
the other of its tapes' heads stationed at some stored record of a long 
prefix of $x$, as ``insurance''.  The early real-time multihead 
simulations of buffers \cite{MRF67, FMR72, Be74} do follow this
strategy, but we show that a machine with only two tapes will not be
able to afford always to use one in this way for insurance:  There will 
have to be a significant subcomputation in which the heads on {\emph 
both\/} tapes ``keep moving'', even ``essentially 
monotonically''---essentially as they would for straightforward 
``copying''.  Under these circumstances, in fact, we will be able to use 
part of the computation itself, rather than the combination of the two 
tapes' contents, as the very redundant representation, to contradict the
following lemma, which we prove later.

     \proclaim{Anti-Holography Lemma}  Consider any constant $C$, and 
consider any binary string $x$ that is long in terms of~$C$, and that is 
nearly incompressible.\footnote{We need $K(x) > \delta|x|$, for some 
fraction $\delta$ that is determined by $C$; so certainly $K(x) > |x| - 
\sqrt{|x|}$ will be enough if $x$ is long.}  Suppose $y = y_1 y_2 \dots 
y_k$ (each $y_i$ a binary string) is a ``representation'' with the 
following properties:
\roster
\item $|y| \le C|x|$;
\item For each $\ell \le k$, $x$'s prefix of length $\ell |x|/k$ is
coded by $y_{i+1} \dots y_{i+\ell}$ for each $i \le k-\ell$.
\endroster
Then $k$ is bounded by some constant that depends only on $C$.
\endproclaim

     For (the binary representation of) a $T$-step subcomputation by $M$ 
to serve as a representation $y$ that {\emph contradicts\/} this lemma, 
we need the following:
\roster
     \item A nearly incompressible input prefix $x$ of length at least
$|y|/C = \varTheta(T/C)$ was read before the subcomputation.
     \item There is a parse of the subcomputation into a large
number~$k$ of pieces so that each prefix of $x$ of length $\ell |x|/k$
is coded in every contiguous sequence of $\ell$ pieces.
     \item $k$ is (too) large in terms of $C$.
\endroster
We accomplish these things by finding a subcomputation that has a
spatially monotonic ``matching'' that is both long and so well separated 
spatially that needed information on tape contents cannot be spread over 
many pieces of the subcomputation.

     The first step is to define and find ``a large matching'', and the 
second is to refine it in a suitable way.  In a two-tape or two-head 
computation or subcomputation, a monotonic sequence of time instants is 
a {\demph matching\/} if neither head scans the same tape square at more 
than one of the time instants.  (So there is actually a separate 
one-to-one ``matching'' for each head, between the time instants and the 
tape squares scanned by that head at those times.)  We prove the 
following lemma later on.

     \proclaim{Large-Matching Lemma}  If a two-tape Turing machine
recognizes
$$\{\, x2x' \mid \text{$x \in \{0,1\}^*$ and $x'$ is a prefix of $x$}\,\}$$
in real time, then its computation on an incompressible binary input of 
length $n$ includes a matching of length $\varOmega(n)$.  (The implicit 
constant does depend on the machine.)\endproclaim

\noindent (Note that this lemma does {\emph not\/} hold if the two heads 
can be on the same tape.)

\smallskip

     In a two-tape or two-head computation or subcomputation, a matching 
is (spatially) {\demph monotonic\/} if, for each of the two heads, the 
spatial order of the corresponding sequence of tape squares being 
scanned at the specified time instants is strictly left-to-right or 
strictly right-to-left. The {\demph minimum separation\/} of a monotonic 
matching is the least distance between successive tape squares in either 
corresponding sequence of tape squares.

     \proclaim{Monotonization Lemma}  Suppose $\varepsilon > 0$ is small
in terms of $\delta > 0$.  If a two-tape (sub)computation of length $T$ 
has a matching of length at least $\delta T$ and internal overlap less 
than $\varepsilon T$, then the computation has a\/ {\rm monotonic
submatching} of length $\varOmega(\delta/\varepsilon)$ and minimum
separation $\varOmega(\varepsilon T)$.  (The implicit constants here
really are constant, not depending even on the machine; for use below, 
let $c$ denote the smaller of them.)\endproclaim

     \demo{Proof}  Without loss of generality, assume $T$ is large in
terms of $\delta$ and $\varepsilon$.  Parse the computation into about
$\delta/(2\varepsilon)$ subcomputations, each including a matching of
length at least $2 \varepsilon T$.  Each subcomputation involves a
contiguous set of at least $2 \varepsilon T$ distinct tape squares on
each tape.  The sets from successive subcomputations touch or intersect, 
but the overlap bound limits their intersection to less than 
$\varepsilon T$ tape squares.  If we omit every second subcomputation's 
set, therefore, we get a spatially monotonic sequence of about 
$\delta/(4\varepsilon)$ {\emph non\/}intersecting sets on each tape.  If 
we further omit every second {\emph remaining\/} set, then we get a 
monotonic sequence of about $\delta/(8\varepsilon)$ sets on each tape, 
with successive sets separated by at least $2 \varepsilon T$ tape 
squares.  To get the desired submatching, simply include one 
matching-time instant from each of the $\delta/(8\varepsilon)$ remaining 
subcomputations.\qed\enddemo

\head 4.  Careful Argument \endhead
     Now let us put together the whole argument, taking care to
introduce the ``constants'' $M$ (and $d$), $\delta$, $\varepsilon$, and
$\varepsilon'$ in an appropriate order, all before the input length~$n$
and the particular input string $x_0$ on which we focus.  Each of these 
values is allowed to depend on earlier ones, but not on later ones.

     For the sake of argument, suppose some two-tape Turing machine $M$ 
does recognize the language $\{\, x2x' \mid \text{$x \in \{0,1\}^*$ and 
$x'$ is a prefix of $x$}\,\}$ in real time, say with delay bound $d$.  
Citing the Large-Matching Lemma, take $\delta > 0$ small enough so that 
$M$'s computation on any incompressible input string $x \in \{0,1\}^*$ 
includes a matching of length at least $\delta |x|$.  Let $\varepsilon > 
0$ be small in terms of $d$, $\delta$, and $M$; and let $\varepsilon'$ 
be small in terms of $d$, $\delta$, and $\varepsilon$.  Let $n$ be large 
in terms of {\emph all\/} these constants, and let $x_0$ be any 
incompressible string of $n$ bits.

     Split the computation by $M$ on input $x_0$ into an initial
subcomputation and a final subcomputation, each including a matching of 
length $\lfloor \delta n/2 \rfloor$.  The number of {\emph steps\/} in 
each of these subcomputations will lie between $\lfloor \delta n/2 
\rfloor$ and $dn$.  Therefore, the initial one will involve a prefix of 
$x_0$ of length at least $(1/d)(\delta n/2) = n \delta/(2d)$, and the 
final one will have ``match density'' at least $(\delta n/2)/(dn) = 
\delta/(2d)$.

     Applying the Overlap Lemma to the final subcomputation above, we
obtain a subcomputation of some length $T \ge \varepsilon'n$, with match
density at least $\delta/(4d)$ and relative internal overlap less than
$\varepsilon$, provided $\varepsilon'$ was chosen small enough in terms 
of $d$, $\delta$, and $\varepsilon$. Then applying the Monotonization 
Lemma, we obtain within this subcomputation a monotonic submatching of
minimum separation at least $c \varepsilon T$, and of length $2k+1$,
where $2k+1$ is either $\lceil c(\delta/(4d))/\varepsilon \rceil$ or
$\lceil c(\delta/(4d))/\varepsilon \rceil-1$ (whichever is odd).  If 
$\varepsilon$ was chosen small, then $k$ will be large.  Note that 
$k\varepsilon$ is approximately equal to a constant $c\delta/(8d)$ that 
depends only on $M$.

     To obtain the desired contradiction to the Anti-Holography Lemma, 
take $y$ to be a complete record of the $T$-step subcomputation obtained 
above, including the symbols scanned and written by each head on each 
step.  To obtain $y_1$, $y_2$, \dots, $y_k$, split this record at every 
second one of the time instants corresponding to the matching of length 
$2k+1$, starting with the third and ending with the third-to-last.  Take 
$x$ to be $x_0$'s prefix of length $kc\varepsilon T/(2d)$.  Since 
$\delta n/(2d)$ exceeds this length (assuming we chose our constants 
appropriately), all of $x$ was already read during the initial 
subcomputation above, and hence before the beginning of the 
subcomputation described by $y$. Note that, for some constant $D$ that 
depends only on $M$,
$$|y| \le DT = \frac{2dD}{kc\varepsilon}|x| \approx \frac{16d^2D}{c^2\delta}|x|,$$
and that $k$ is large (in fact, {\emph too\/} large for the
Anti-Holography Lemma) in terms of the constant $C =
16d^2D/(c^2\delta)$, assuming we chose $\varepsilon$ small enough.

     To see that $x$'s prefix of length $\ell|x|/k$ is coded by $y_{i+1} 
\dots y_{i+\ell}$ (for each appropriate $\ell$ and $i$), suppose we 
interrupt $M$ with ``the command to begin retrieval'' (i.e., with the 
symbol $2$) at the $(2i+\ell+1)$st of the time instants corresponding to 
the matching of length $2k+1$.  Since $M$ must be able to check the 
prefix of length $\ell|x|/k$ by reading only the information within 
distance $d\ell|x|/k = \ell c \varepsilon T/2$ of its heads, that prefix 
must be coded by that information.  Since this distance in each 
direction is just $\ell/2$ times the minimum separation of the matching, 
and since the matching is monotonic, the same information is available 
within the subcomputation record~$y$, between the matching's time 
instants $2i+\ell+1-\lceil\ell/2\rceil$ and 
$2i+\ell+1+\lceil\ell/2\rceil$.  Since $y_{i+1} \dots y_{i+\ell}$ runs 
from the matching's time instant $2i+1 \le 2i+\ell+1-\lceil\ell/2\rceil$ 
to the matching's time instant $2i+2\ell+1 \ge 
2i+\ell+1+\lceil\ell/2\rceil$, it too codes the desired prefix.

\head 5.  Proof of Anti-Holography Lemma \endhead
     Without loss of generality, assume $k$ is equal to $2^e$ for some 
integer exponent~$e$.\footnote{If it is {\emph not}, then just reduce it 
until it {\emph is}.} Then the target constant can be $2^{2C-1}$. Again 
without loss of generality, assume $k$ is {\emph at most\/} this target 
constant {\emph times two}.\footnote{Otherwise, pair up $y_i$'s to 
reduce $k$ by factors of $2$ until it {\emph is}.} Finally, without loss 
of generality, assume that $|x|=n$ is divisible by $k$, with $x = x_1 
\dots x_k$ and $|x_i| = n/k$ for every~$i$.\footnote{If $x$'s length is 
{\emph not\/} divisible by $k$, then just discard at most its last 
$2^{2C-1}$ bits, until its length {\emph is\/} divisible by $k$.}

     To obtain short descriptions of $y$, we abbreviate many of its
subwords in terms of just a few prefixes of $x$, using the symmetry of 
information.  For each $j \le e$, and for $j=e-1$ in particular, this 
will yield
$$K(y | x_1 \dots x_{2^j}) \le |y| - (1+j/2) n + \BO(\log n).$$
Unless $k$ is smaller than $2^{2C-1}$, $e-1$ will be so large that this 
will imply that $x_1 \dots x_{2^{e-1}}$ codes $y$.  Since $y$ in turn 
codes all of $x = x_1 \dots x_{2^e}$ this will mean that the first half 
of $x$ codes the whole string, contradicting the incompressibility 
assumption for $x$.

     By induction on $j$ ($j = 0$, $1$, \dots, $e$), we actually prove 
``more local'' bounds that {\emph imply\/} the ones above:  For each 
appropriate $i$ ($i = 0$, $1$, \dots, $k-2^j$),
$$K(y_{i+1} \dots y_{i+2^j} | x_1 \dots x_{2^j})
\le |y_{i+1} \dots y_{i+2^j}| - 2^j (1+j/2) n/k + \BO(\log n).$$
Both the base case and the induction step are applications of an
intuitively clear corollary, the Further-Abbreviation Lemma below, of
the symmetry of information.  For the base case, we apply the lemma with 
$y'$ equal to~$y_{i+1}$, $x'$ equal to the null string, and $x''$ equal 
to~$x_1$, to get the desired bound on $K(y'|x'')$:
$$\align
K(y'|x'') &\le K(y') - K(x'') + \BO(\log n)\\
          &\le |y'| - n/k + \BO(\log n).
\endalign$$
For the induction step, we let $y'' = y_{i+1} \dots y_{i+2^j}$ and $y''' 
= y_{i+2^j+1} \dots y_{i+2^{j+1}}$, and apply the lemma with $y'$ equal 
to $y''y'''$, $x'$ equal to $x_1 \dots x_{2^j}$, and $x''$ equal to 
$x_{2^j + 1} \dots x_{2^{j+1}}$, to get the desired bound on
$K(y'|x'x'')$:
$$\align
K(y'|x'x'') &\le K(y'|x') - K(x'') + \BO(\log n)\\
 &\le K(y'' | x') + K(y''' | x') - K(x'') + \BO(\log n)\\
 &\le |y''| + |y'''| - 2 \cdot 2^j (1+j/2) n/k - 2^j n/k + \BO(\log n)\\
 &=   |y''| + |y'''| - 2^{j+1} (1+(j+1)/2) n/k + \BO(\log n).\quad\qed
\endalign$$

     \proclaim{Further-Abbreviation Lemma} Assume $y'$, $x'$, and $x''$ 
are strings of length $\varTheta(n)$, with
$$K(x''|y') = \BO(\log n)$$
and
$$K(x''|x') = K(x'') - \BO(\log n).$$
(I.e., $y'$ codes $x''$, which is nearly incompressible relative to
$x'$.)  Then
$$K(y'|x'x'') \le K(y'|x') - K(x'') + \BO(\log n).$$\endproclaim

     \demo{Proof} Let $d(u|v)$ denote a shortest description of $u$ in
terms of $v$, so that $\bigl|d(u|v)\bigr| = K(u|v)$.  Then
$$\align
K(y'|x'x'') &\le K(d(y'|x')|x'x'') + \BO(\log n)\\
 &\le K(d(y'|x')|x'' \bigm| x') + \BO(\log n)\\
 &\le K(d(y'|x') \bigm| x') - K(x'' \bigm| x') + K(x''|d(y'|x') \bigm| x') + \BO(\log n)\\
 &\le K(y'|x') - K(x''|x') + K(x''|y') + \BO(\log n)\\
 &\le K(y'|x') - K(x'') + \BO(\log n).\qed
\endalign$$\enddemo

\head 6.  Proof of Large-Matching Lemma \endhead
     Our proof of the Large-Matching Lemma is based on an earlier
theorem of Vit\'anyi:

     \proclaim{Far-Out Lemma \cite{Vi84}\footnote{For a {\emph sketch\/} 
of the proof, see the appendix below.}}  If a two-tape Turing machine 
recognizes
$$\{\, x2x' \mid \text{$x \in \{0,1\}^*$ and $x'$ is a prefix of $x$}\,\}$$
in real time, then its ``worst-case closest head position''\footnote{If 
$p_i(t)$ denotes the net displacement of head $i$ at time $t$, then the 
``worst-case closest head position'' is $\max_t \min_i p_i(t)$.} on 
incompressible inputs $x \in \{0,1\}^n$ is $\varOmega(n)$.\endproclaim

\noindent In other words, incompressible binary data is guaranteed at
some point to drive {\emph both\/} heads of such a machine {\emph
simultaneously\/} far from their original positions.  By the continuity 
of sequential access, of course, this means that the heads actually 
spend {\emph long intervals\/} of time simultaneously far from their 
original positions; and this is the fact that we exploit.

     We actually show that even any two-{\emph head\/} Turing machine
(with both heads on the {\emph same\/} one-dimensional tape) that
recognizes our language and that satisfies the conclusion of the Far-Out 
Lemma also satisfies the desired conclusion of the Large-Matching Lemma. 
(Of course the obvious two-head machine, that does recognize our language 
in real time, does not satisfy the conclusion of {\emph either\/} lemma.) 
This simplifies the exposition, since we have only one tape to talk 
about.  Note that the ``matching'' notion does make sense even when both 
heads are on the same tape.

     As earlier, let us take explicit care to introduce our
``constants'' in an appropriate order.  Consider any two-head Turing
machine $M$ alleged to recognize
$$\{\,x2x' \mid \text{$x \in \{0,1\}^*$ and $x'$ is a prefix of $x$}\,\}$$
in real time, say with delay bound $d$, and that satisfies the
conclusion of the Far-Out Lemma.  Let $c$ be small enough to serve as
the implicit constant in that conclusion.  Let $\varepsilon$ be small in
terms of $M$ and $c$; let $\delta$ be small in terms of $M$, $c$, and
$\varepsilon$; let $n$ be large in terms of $M$, $c$, $\varepsilon$, and
$\delta$; and let $x$~be an incompressible string of $n$ bits. 
Exploiting the conclusion of the Far-Out Lemma, parse $x$ into three
pieces, $x=uvw$, such that $uv$ leaves both heads at least $cn$ tape
squares from where they started and the length of $u$ is $\lfloor
cn/(3d) \rfloor = \varTheta(n)$.

     Consider $M$'s computation on $uv2u$.  The first $u$ must be read
before either head gets as far as even $cn/3$ tape squares from where it 
started, but the second~$u$ must be read while neither head gets {\emph 
closer\/} than $2cn/3$ tape squares to where it started. During its 
subcomputation on $v$, therefore, it seems that $M$ must somehow 
``copy'' its representation of $u$ across the intervening $cn/3$ tape 
squares.  We show that this process has to involve a matching larger 
than $\delta n$.

     For the sake of argument, suppose there is {\emph not\/} a matching 
larger than $\delta n$.  Then there must be a {\emph maximal\/} matching 
of size only $m \le \delta n$.  We will select some correspondingly 
small ``interface'' through which a description of $u$ must pass.  That 
interface will involve some rarely crossed boundary at distance between 
$cn/3$ and $2cn/3$ from the heads' starting position, and some other 
rarely crossed boundaries that tightly isolate the $2m$ tape squares 
involved in the matching.  Since there are $2cn/3-cn/3$ candidates for 
the former, we can select one that is crossed only a constant number 
(bounded in terms of $d$ and $c$) of times.  We will refer to the tape 
squares on the respective sides of this selected {\demph central 
boundary\/} as {\demph close\/} and {\demph far}.  By the following 
purely combinatorial lemma, we can tightly isolate the matched tape 
squares with at most $4m$ additional boundaries, each of which is 
crossed only a constant number (bounded in terms of $d$, $c$, and our 
``tightness criterion'' $\varepsilon$) of times.

     \proclaim{Tight-Isolation Lemma} Consider a finite sequence $S$ of 
nonnegative numbers, the first and last of which are $0$.  Let some of 
the separating ``commas'' be specially designated---call them
``semicolons''.  For each threshold $\ell \ge 0$, let $S_\ell$ be the
subsequence consisting of the items\footnote{Note that the number of
such {\emph items\/} can be small even if the number of semicolons is
large.  For $\ell$ large enough, in fact, $|S_\ell|$ will be $0$.} that 
are reachable from the semicolons via items that exceed $\ell$ (and that 
themselves exceed $\ell$).  Then, for each $\varepsilon > 0$, there is 
some $\ell$ such that $\ell|S_\ell| < \varepsilon \sum S$, where $\sum 
S$ denotes the sum of the entire sequence $S$ and $\ell$ is bounded by 
some constant that depends only on $\varepsilon$.\endproclaim

     \demo{Proof} Let $T = \sum S$, and let $k = 2 \lceil 2/\varepsilon
\rceil$.  Since $2T/k \le \varepsilon T/2 < \varepsilon T$, let us aim 
for $\ell|S_\ell| \le 2T/k$.  If no $\ell$ in $\{\,k^i \mid 0 \le i \le
k\,\}$ were to work, then we would have
$$2T/k < k^i|S_{k^i}| < T$$
for every $i$.  But this would lead to the contradiction
$$\align T &> \sum_{i=0}^k k^i(|S_{k^i}|-|S_{k^{i+1}}|)\\
           &> \sum_{i=0}^k (2T/k - T/k)\\
           &= (k+1)T/k.\quad\qed
\endalign$$\enddemo

\noindent In our application, the numbers are the lengths of the
crossing sequences associated with the boundaries between tape squares, 
their sum is at most $dn$, and the semicolons are the matched tape 
squares.  We obtain our desired ``isolation neighborhoods'' from the 
at-most-$2m$ contiguous neighborhoods that comprise $S_\ell$\footnote{To 
include all the semicolons, some of these ``contiguous neighborhoods'' 
might have to be the {\emph empty\/} neighborhoods of the semicolons.} 
by adding one item at each end of each neighborhood.  (This might cause 
neighborhoods to combine.)  This adds at most $4m$ items to $S_\ell$ and 
results in nonempty isolation neighborhoods whose boundary items are at 
most $\ell$.

     Actually, the picture is clearer if we select our central boundary 
{\emph after\/} we select the isolation neighborhoods. Assuming 
$\varepsilon$ and $\delta$ are chosen appropriately small, this lets us 
select a boundary not included in any of the isolation neighborhoods.  
(There are at most $4m + |S_\ell| \le 2\delta n + \varepsilon dn \le 
cn/6$ boundaries (half the original number of candidates) to avoid.)

     Finally, we use our suggested interface to give a description of
$u$ in terms of $v$ that is too short---say shorter than $|u|/2 \approx 
cn/(6d)$.  (We could substitute a description this short for $u$ in $x$ 
to contradict the incompressiblity of $x$.)  We claim we can reconstruct 
$u$ from $M$, $v$, the length of $u$, and the following information 
about the subcomputation of $M$ while reading the $v$ part of input 
$uv$:
\roster

     \item The sequence of all $\BO(m)$ selected boundary locations.

     \item The sequence of all $\BO(m)$ crossings of these selected
boundaries, and their times (implicitly or explicitly including the
corresponding input positions).

     \item The following information for each close-to-far crossing, and 
for the end of the subcomputation:

     \itemitem"{$\bullet$}" $M$'s control state and head positions.

     \itemitem"{$\bullet$}" The full content of every isolation
neighborhood.

     \item The following information for each crossing {\emph out of\/} 
an isolation neighborhood:

     \itemitem"{$\bullet$}" The full content of that isolation
neighborhood.

     \itemitem"{$\bullet$}" The full content of the isolation 
neighborhood in which the other head remains\footnote{The other head 
must remain in {\emph some\/} isolation neighborhood---otherwise, the 
matching could be enlarged.}---{\emph provided\/} that there has been a 
new crossing into that neighborhood since the previous time such 
information was given for it.

\endroster

     To determine $u$, it suffices to reconstruct enough of $M$'s
configuration after its computation on input $uv$ so that we can check
which additional input string $2u'$ of length $1+|u|$ leads to
acceptance.  The far tape contents suffice for this.

     Our reconstruction strategy is mostly to simulate $M$ step-by-step, 
starting with the first close-to-far crossing.  Toward this end, we 
strive to maintain the contents of any currently scanned close isolation 
neighborhood and of the entire far side.  We temporarily {\emph 
suspend\/} step-by-step simulation whenever a head shifts onto a {\emph 
close\/} tape square not in any isolation neighborhood, and we aim to 
{\emph resume\/} suspended step-by-step simulation whenever a head 
shifts onto a {\emph far\/} tape square not in any isolation 
neighborhood.  Because our matching is maximal, such a far tape square 
is {\emph not\/} scanned at the time of suspension, and hence also not 
at any time before the desired resumption.  It follows that the 
information for the needed updates is indeed available, so that 
resumption is indeed possible.  Similarly, any necessary updates are 
possible if the step-by-step simulation happens to be suspended when the 
subcomputation ends.

     It remains only to show that $|u|/2$ bits suffice for our
description of $u$ in terms of~$v$.  For each of the sequences in (1)
and (2), the trick is to give only the first number explicitly, and then 
to give the sequence of successive differences.  The length of this 
encoding is $\BO(m\log(n/m)) = \BO(n\log(1/\delta)/(1/\delta))$, which 
can be limited to a small fraction of $|u|/2 \approx cn/(6d)$ by 
choosing $\delta$ small enough.  For (4), note that that the contents of 
each isolation neighborhood is given at most once for each of the $\ell$ 
crossings into and out of the neighborhood.  For (3) and (4), therefore, 
straightforward encoding requires only $\BO(\log n + \ell(m+|S_\ell|)) = 
\BO(\log n + \ell\delta n + \varepsilon dn)$ bits, where the implicit 
constant is bounded in terms of $d$ and $c$.  This can be limited to 
another small fraction of $|u|/2$ by choosing $\varepsilon$ small 
enough, $\delta$ small enough, and $n$ large enough.  For the remaining 
information, $M$, $|u|$, and a description of this whole discussion, we 
need only $\BO(\log n)$ bits, which can be limited to a final small 
fraction of $|u|/2$ by choosing $n$ large enough.\qed

\head 7.  Further Discussion and Remaining Questions \endhead
     In retrospect, our contribution has been a constraint on how a
Turing machine with only two storage heads can recognize $L$ in real
time.  Even if the two heads are on the {\emph same\/} one-dimensional 
tape, such a Turing machine cannot recognize $L$ in real time unless it
violates the conclusion of (the first sublemma of) Vit\'anyi's Far-Out 
Lemma (see Appendix below).  Only in the latter do we ever really 
exploit an assumption that the two heads are on separate tapes.

     Our result rules out general real-time simulation of a two-head
tape unit using only a pair of single-head tapes.  It remains to be
investigated whether the result extends to some notion of {\emph 
probabilistic\/} real-time simulation \cite{{\rm cf.,} PSSN90}.  Another 
extension might rule out simulation using {\emph three\/} single-head 
tapes, yielding a tight result; but this would require a more difficult 
witness language. Perhaps allowing the ``back'' head of the defining 
two-head machine also to move and store random data, but much more 
slowly than the ``front'' head, would let us combine our arguments with 
those of Aanderaa \cite{Aa74, PSS81, Pa82}.  A slightly weaker 
possibility might be to show that {\emph two\/} single-head tapes {\emph 
and a pushdown store\/} do not suffice, and a slightly stronger one 
might be to show that even {\emph three\/} single-head tapes and a 
pushdown store do not suffice.

     It might be even more difficult to rule out general real-time
simulation of a two-head one-dimensional tape unit using two or three
{\emph higher-dimensional\/} single-head tapes.  Our particular language
$L$ {\emph can\/} be recognized in real time by a Turing machine with
just two such two-dimensional tapes---the idea is to strive to maintain 
the $n$ bits of data within an $\BO(\sqrt n)$ radius on both tapes, along 
with $\BO(\sqrt n)$ strategically placed {\emph copies\/} of the first 
$\BO(\sqrt n)$ bits, to serve as insurance {\emph alternatives\/} at the 
same time that the array of their left ends provides a convenient area 
for temporary collection of data and for copying data between the tapes.

     The implications for real-time simulation of one-dimensional tape
units with {\emph more than\/} two heads remain to be investigated. For 
example, how does a three-head tape compare with three single-head tapes 
or with one single-head tape and one two-head tape?  (Paul's results 
\cite{Pa84} do answer such questions for tapes of higher dimension.)  
How tight is the known bound of $4h-4$ single-head tapes for real-time 
simulation of one $h$-head (one-dimensional) tape \cite{LS81}? Perhaps 
the many-heads setting is the right one for a first proof that even an 
{\emph extra\/} head is not enough to compensate for the loss of 
sharing; e.g., can a 1000-head tape be simulated in real time by 1001 
single-head tapes, or by 1000 single-head tapes and a pushdown store?

     Finally, does any of this lead to more general insight into the
heads or tapes requirements for arbitrary computational tasks?  I.e.,
when asked about some computational task, can we tightly estimate the
structure of the sequential storage that suffices for the task?

\head  Appendix:  A proof sketch for Vit\'anyi's Far-Out Lemma \endhead
     Suppose two-tape Turing machine $M$ recognizes the language in real 
time.  Without loss of generality, assume $M$'s storage tape is only 
{\emph semi\/}-infinite, and assume $M$ writes only $0$'s and $1$'s.  
Let $d$ be the delay of $M$.

     Our ultimate goal is to show that both heads {\emph
simultaneously\/} range linearly far when the input is incompressible,
but first we show that each one {\emph separately\/} does so even when
the input is just {\emph nearly\/} incompressible.  (The subsequent
application is to significantly long {\emph prefixes\/} of input strings
that are not compressible at all.)  It is only this part of the proof
that requires the hypothesis that the two heads are on separate tapes.
This part is based on the ``bottleneck'' argument that Valiev 
\cite{Va70} (and, independently, Meyer \cite{Me71}) used to show that no 
{\emph single\/}-tape Turing machine can accept the simpler language 
$\{\, x2x \mid x \in \{0,1\}^* \,\}$ in real time.

     Suppose $\varepsilon$ is small in terms of $M$ and $d$, $n$ is large 
in terms of all of the above, and $x$ is of length $n$ and nearly 
incompressible ($K(x) \ge n - \sqrt n$).  We want to show that each head 
ranges farther than $\varepsilon n$.

     Suppose the head on one of the tapes, say the first, does {\emph 
not\/} range farther than $\varepsilon n$.  Then the head on the {\emph 
second\/} tape must certainly range farther than, say, $n/3$.  
(Otherwise, the total state after storage of $x$ is a too-short 
description of $x$.)  Let $uvw$ be the parse of $x$ with $uv$ the 
shortest prefix of $x$ that leaves $M$'s second head at least $n/3$ tape 
squares out, and with $|u| = n/(9d)$, so that that same head gets no 
farther than $n/9$ tape squares out during input of $u$.  On that head's 
tape, there must be a ``bottleneck'' boundary between $n/9$ and $2n/9$ 
tape squares out that gets crossed at most $9d$ times.  Since all of $u$ 
gets read when the second head is to the left of this bottleneck, it is 
possible to describe $x = uvw$ in terms of $vw$ and the bottleneck's 
``crossing sequence'', which should include, for each crossing, the step 
number and the ``crossing state'', which in turn should include the 
complete but relatively small contents of the {\emph first\/} storage 
tape at the time of the crossing. The following information suffices:
\roster

     \item $vw$,

     \item a description of this discussion,

     \item a description of $M$,

     \item the value of $n$,

     \item the location of the bottleneck,

     \item the crossing sequence at the bottleneck.

\endroster
If we provide $vw$ as a literal suffix, then we can limit the length of 
this description to little more than $n - |u|$ bits, contradicting the 
near incompressibility of $x$.  To recover $u$, we can use the 
information to determine enough of $M$'s instantaneous description after 
reading $uv$ (omitting from the i.~d.~only what is to the left of the 
bottleneck on the second tape) to then try each input continuation $2u'$ 
with $|u'| = n/(9d)$.

     Finally, we return to our ultimate goal.  Here is the idea:  If the 
heads do {\emph not\/} both go far out together, then they must {\emph 
take turns}, so that some region gets crossed many times; abbreviate the 
symbols read while a head is in that region.

     Suppose $\varepsilon$ is small in terms of $M$ and $d$ (as above), 
$\varepsilon_2$ is small in terms of the preceding parameters (in 
particular, $\varepsilon_2 \ll \varepsilon$), $\varepsilon_1$ is small in 
terms of the now preceding parameters (in particular, $\varepsilon_1 \ll 
\varepsilon_2$), $n$ is large in terms of all of the above, and $x$ is of 
length $n$ and incompressible.  We want to show that both heads range 
farther than $\varepsilon_1 n$, simultaneously.

     Suppose, to the contrary, that there is always at least one head 
within $\varepsilon_1 n$ tape squares of the origin.  We count the 
crossings of the region from $\varepsilon_1 n$ to $\varepsilon_2 n$:  It 
follows from our assumptions that a left-to-right crossing must occur 
between input symbol number $(d/\varepsilon)^i (\varepsilon_2 
n/\varepsilon)$ and input symbol number $(d/\varepsilon)^{i+1} 
(\varepsilon_2 n/\varepsilon)$, for every~$i$.  (We use the fact that 
these input prefixes are themselves nearly incompressible.)  By input 
symbol number $n$, therefore, the number of complete crossings (either 
direction) is at least $r = 2 \log_{d/\varepsilon} 
(\varepsilon/\varepsilon_2)$ (which is large because $\varepsilon_2$ is 
so small).

     There is a complication, however:  There might also be {\emph 
partial\/} crossings, involving fewer input symbols but additional 
overhead in the description we plan to give.  To control this problem, we 
shrink the region slightly, replacing $\varepsilon_1$ and $\varepsilon_2$ 
with $\varepsilon_1'$ and~$\varepsilon_2'$ from the first and last 
quarters, respectively, of the range $[\varepsilon_1,\varepsilon_2]$, 
chosen so that each of the boundaries $\varepsilon_1' n$ and 
$\varepsilon_2' n$ is crossed at most $R = 8d/\varepsilon_2$ times.  
This is possible, since $R (\varepsilon_2 - \varepsilon_1) n/4$ exceeds 
$dn$.

     Finally, then, we formulate a description of the incompressible 
input that differs from the completely literal one as follows:  We {\emph 
eliminate\/} the input read while a head is in the range between 
$\varepsilon_1' n$ and $\varepsilon_2' n$, for a {\emph savings\/} of at 
least $r(\varepsilon_2' - \varepsilon_1') n/d \ge r(\varepsilon_2 - 
\varepsilon_1) n/(2d)$ bits.  We {\emph add\/} descriptions of the 
crossing sequences at these two boundaries, including times, states, and 
the tape contents out to boundary $\varepsilon_1 n$, and also the full 
{\emph final\/} contents of the tape squares {\emph between\/} the two 
boundaries, for a total {\emph cost\/} of
$$\BO((\varepsilon_2'-\varepsilon_1')n + R (\log n + \varepsilon_1 n))
= \BO((\varepsilon_2-\varepsilon_1)n + 8d (\log n + \varepsilon_1 n)/\varepsilon_2)
=\BO((\varepsilon_2-\varepsilon_1)n)$$
bits, which can be kept significantly smaller than the savings.\qed

\enddocument